%% file: Kiyaeva_n.tex
\documentclass[
 aps, pra,
 amsmath,amssymb,
 11pt,
 final,
tightenlines,
 twoside,
 twocolumn,
 nofloats,
nofootinbib,
 superscriptaddress,
showkeys,
showkeywords,
 ]
{revtex4-2}

\usepackage[T1]{fontenc}
\usepackage[utf8x]{inputenc}
\usepackage[english]{babel}
\usepackage{graphicx}
\usepackage{dcolumn}
\usepackage{bm}

\input{maik.rty}

\setcitestyle{authoryear,round}
\input{sao_cmd_author.tex}

\begin{document}

\selectlanguage{english}



\title{A Promising New Dark Matter Candidate, and Implications for the Search for Extraterrestrial Intelligence (SETI)}

\author{\firstname{A.}~\surname{Prillfool}}
 \email{apf0104@cedds.cam.ac.uk}
 \affiliation{St. Cedd's College, University of Cambridge,}

\author{\firstname{A. A.}~\surname{Stoffers}}
 \affiliation{Kavli Institute of Cosmology, University of Cambridge, Madingley Road, Cambridge CB3 0HA, UK}

\author{\firstname{I.}~\surname{Juodžbalis}}
 \affiliation{Kavli Institute of Cosmology, University of Cambridge, Madingley Road, Cambridge CB3 0HA, UK}

\author{\firstname{M. S.}~\surname{Bothwell}}
 \affiliation{Institute of Astronomy, University of Cambridge, Madingley Road, Cambridge CB3 0HA, UK}

 \author{\firstname{A. J.}~\surname{Wojcik}}
 \affiliation{Electrical Engineering Division, Department of Engineering, University of Cambridge, 9 JJ Thomson Avenue, Cambridge CB3 0FA, UK}

\begin{abstract}

\noindent We present a speculative exploration of the properties of a proposed new Dark Matter (DM) candidate in a heretofore under-explored region of parameter space. Our proposed ultra-cold candidatae has been a matter of speculation for some time,and has recently been tentatively identified via direct-detection. While unconventional, demonstrated existence of this DM candidate would have wide-ranging implications for a range of fields, from particle cosmology to exobiology and the Search for Extraterrestrial Life (SETI).

\end{abstract}

\maketitle

\section{INTRODUCTION: Dark Matter - A failed search}
For nearly a century, the dark matter problem has been a fundamental issue in astrophysics, bearing a striking resemblance to the historical challenge of detecting unseen planets. Anomalies in large astrophysical systems can be interpreted in two ways: as either challenges to the established laws of gravitation and general relativity or as evidence of invisible, dark objects. Initially, it took decades for the scientific community to recognize dark matter as a genuine issue. Today, comprehensive data from the study of microwave background radiation, supernova distance measurements, and large-scale galaxy surveys have reinforced the standard model of cosmology. This model suggests that the universe's structure emerged from the gravitational amplification of minor density fluctuations, facilitated by cold dark matter (CDM). The presence of dark matter is crucial; without it, the density contrasts observed in the universe today could not have evolved from the modest amplitude of density fluctuations seen in the cosmic microwave background.\\
Candidates for dark matter vary widely, spanning 90 orders of magnitude in mass, from ultralight bosons to massive primordial black holes. Interest in the latter has surged following the detection of gravitational waves from black hole mergers by the LIGO and Virgo observatories \cite{bird2016did,clesse2017detecting}. For many years, Weakly Interacting Massive Particles (WIMPs) were considered ideal candidates for dark matter. They fit well within the weak-interaction mass scale (approximately 10 GeV to 1 TeV) and could naturally emerge with the correct relic abundance in the early universe \cite{bertone2005particle}. Moreover, WIMPs offered a potential solution to the hierarchy problem \cite{de2014criteria}, a major focus of particle physics for the last four decades. However, despite extensive searches, WIMPs have not been detected.\\
Supersymmetry (SUSY) proposes a partner particle for every particle in the standard model, with the partners of electroweak bosons being potential WIMP candidates and thus suitable for dark matter. However, most of the parameter space of natural simple SUSY models have been largely discounted, leading to questions about the extent of fine-tuning we are willing to accept in pursuit of discovering SUSY \citep{ross2017revisiting}.

While primordial black holes have been considered as dark matter candidates, stringent constraints from various observational probes, such as microlensing and the cosmic microwave background, have largely ruled out their viability over broad mass ranges, casting doubt on their prevalence in the universe \cite{sasaki2018primordial}. Similarly, while axions emerged as promising candidates due to their potential to solve the strong CP problem, recent experimental bounds and theoretical considerations have narrowed the allowed parameter space, challenging their capacity to constitute the entirety of dark matter \cite{irastorza2018new}.
As existing dark matter candidates face increasing scrutiny, there is a growing impetus to explore alternative theoretical frameworks and experimental avenues in the quest for novel dark matter candidates that can better accommodate observational constraints. Next, we will introduce an alternative dark matter candidate in Sec.~\ref{sec:alter} and explore its potential implications in Sec.~\ref{sec:impl}.


\begin{figure*}
\includegraphics[width=\linewidth]{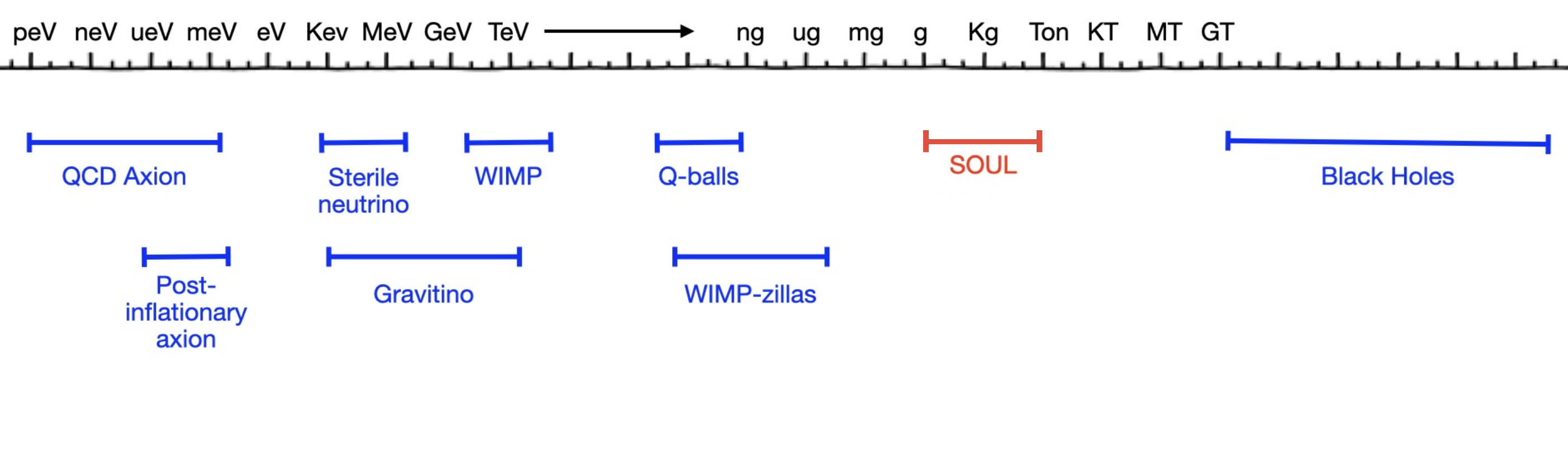}
\caption{A range of proposed dark matter candidates, with their expected mass scale. It can readily be seen that SOUL occupies a yet unexplored area of the mass range.}
\label{fig_DM_part}
\end{figure*}

\section{An alternative DM candidate}
\label{sec:alter}

Several authors have attempted to ascertain the mass of the human soul, most famously MacDougall (1907), who compared pre- and post-mortem mass measurements of the human body, reporting a value of $1.2 \times 10^{28}$~MeV/$c^2$ (=21 grams; an estimate of the error on this value was not reported by the original author). Though the specific methodology employed has been called into question, it is undoubtedly true that throughout the 20th Century a range of experimental and theoretical efforts have attempted to improve this value, with the most recent estimates, utilizing self-consistent treatment of neuron binding energies and brain masses placing the estimate at a much higher value of $10^{30}$~GeV/$c^2$ \citep{soul_weight}. A likely reason for this wide range of estimated masses is the inherent difficulty of performing direct measurements on this elusive substance as its coupling to baryonic matter must be very weak, otherwise its existence would have been well established long ago (given that every human being, presumably, possesses a soul). This means that the soul has two of the fundamental properties of DM -- its might mass and its relative undetectability -- and thus makes for a promising candidate.

This is particularly true in light of the fact that the vast majority of standard DM candidates are in tension with observations of Spontaneous Human Combustion (SHC), as discussed by \cite{DM_SHC} (notably, this paper was published exactly one year ago today). Using SHC \cite{DM_SHC} find an estimated DM particle mass of $5.6 \times 10^{27}$~GeV/$c^2$, which falls comfortably within the mass range discussed above. 

Of course, it is important to avoid anthropocentrism -- it is exceedingly unlikely that specifically \textit{human} souls comprise a significant fraction of the total DM content of the Universe. It is far more likely that other similarly-ensouled beings represent the bulk of the missing matter. We therefore avoid the anthropocentric (and indeed Christocentric) term `soul'. We instead propose the existence of a Sentient/Organic Universal Life-force field (which we will henceforth refer to as a  `SOUL'), with the corresponding \textit{soulon} particle of $10^{25} < M_S < 10^{30}$~GeV/$c^2$, which is compared to other proposed DM candidates in Figure \ref{fig_DM_part}. As can be seen there, the SOUL occupies a place in the parameter space not covered by other models, further cementing its potential as a viable DM candidate.

\section{IMPLICATIONS OF THE SOUL}
\label{sec:impl}

\begin{figure}
    \centering
    \includegraphics[width = \linewidth]{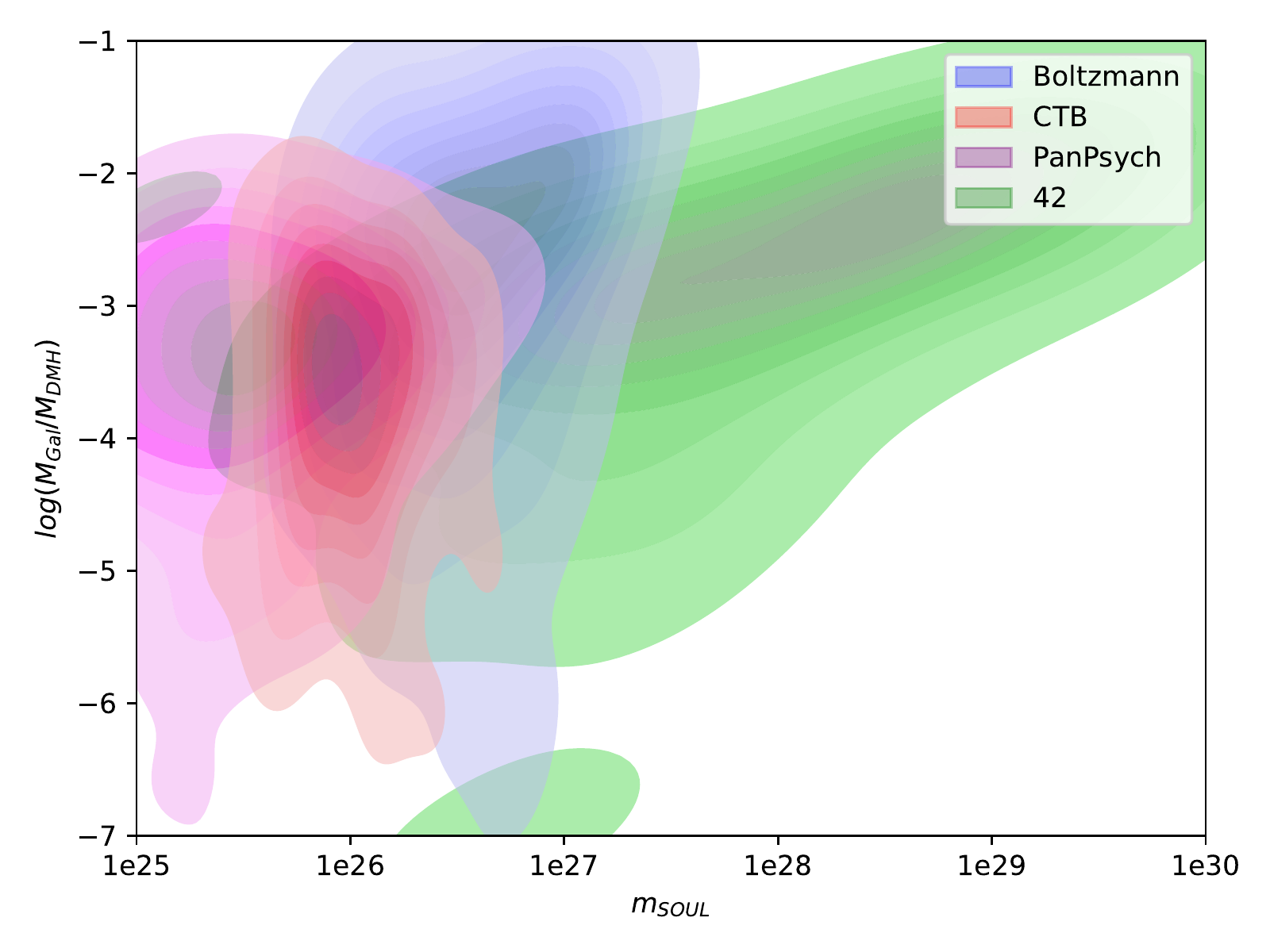}
    \caption{Illustrative figure showcasing the proposed ratio of galaxy mass to dark matter halo mass plotted against probable masses of the SOUL. The shaded regions represent diverse interpretations of observational data and theoretical predictions, offering insights into the potential relationship between galaxies and dark matter. Each shading delineates distinct hypotheses and models.}
    \label{fig:soulDistribution}
\end{figure}

In order to examine the implications of our DM model we first perform a simple order of magnitude estimation in order to quantify the number of SOULs in the Milky Way galaxy.

Taking the DM halo mass of the Milky way to be of order $10^{69}$~GeV/$c^2$ \citep{MW_DM_mass}, we find that the number of SOULs in the galaxy can range between $10^{39} \leq N_S \leq 10^{44}$, depending on the exact model assumed. This greatly outnumbers the number of stars in the Milky Way, which can be estimated to be on the order of $10^{11}$ \citep{MW_stellar_mass}. We estimate the density of SOULs throughout the Milky Way to be approximately $10^{28} \leq N_S \leq 10^{33}$ per star system. 

It is clear, therefore, the number of SOULs required to explain the DM halo of our galaxy likely can not be supplied by intelligent civilizations similar to ours. Assuming each civilization to be similar to our own in terms of the number of conscious individuals -- approximately $10^{11}$ humans have ever lived --  even the extreme situation of one intelligent civilization per star system could explain only a tiny ($10^{-32}$ to $10^{-27}$) fraction of the total Milky Way DM halo mass.

While some might consider this discrepancy a fatal strike against our theory, we prefer to interpret this tension as potential evidence for new physics. Throughout the rest of this paper we further consider the possible interpretations and implications of our model, particularly where the distribution and the true origin of SOULs is concerned. In the following sections we examine several likely scenarios, colourfully illustrated in Figure \ref{fig:soulDistribution}, many of which have been investigated since the dawn of human history. Until now however, they could not be properly evaluated due to a paucity of empirical evidence.

\subsection{The Boltzmann interpretation}
The concept of the Boltzmann brain is used to capture the vastness of possible arrangements of matter allowed by the laws of thermodynamics. Given a sufficiently long time, structures of any complexity are allowed to form, including fully functional human brains. We further expand on this philosophical tool, showing that such structures can occur much more prominently. According to our model, human souls are formed of a large quantity of bound soulons, which we call soulon clouds. It is possible therefore, that such clouds may also form spontaneously in space due to binding of large quantities of free soulons. We call such structures ANGELS (Aggregated Nexus for Global Exchange of Life-force and Soulons). Since soulons allow for transferring consciousness and participate in the thought process, it is possible that the free ANGELS could act similarly to the Boltzmann brains, experiencing dream-like thought and hallucinations. Such ANGELS float through, and exchange particles with, the underlying field of free soulons mentioned earlier, a possible origin of which is further explored in the next section.

\subsection{Cosmic Theological Background scenario}
Another intriguing possibility is that the SOULs unaccounted for by the presence of intelligent life may constitute a single separate intelligence. This idea was first put forward by Giordano Bruno in the 16th century \citep{Bruno1580, Bruno1580b}. These seminal works were among the first to introduce the idea of extraterrestrial life and also brought to bear the idea that, what was defined as 'God' at the time, was uniformly distributed across all space rather than confined to a particular location. The latter idea ended rather poorly for the author as the mathematical and experimental machinery at the time was inadequate to robustly establish it. However, it now presents an intriguing framework within which our results can be interpreted.

To proceed, we shall consider the substance proposed by Bruno as a kind of scalar field with soulons corresponding to its excitations. Such a field would undoubtedly permeate the entirety of the observable Universe, constituting a Cosmic Theological Background (CTB). The CTB could serve as the origin of the free soulons discussed in the previous section and contribute the vast majority of observed DM mass in the Universe. However, the precise treatment of the dynamics and vacuum fluctuations of CTB is beyond the scope of this work, particularly as there are further interpretations of the SOUL possible.

\subsection{The panpsychist interpretation}

An alternate, intriguing possibility is the so-called `panpsychist interpretation' -- that consciousness, rather than being an emergent property of matter arranged to a specific degree of complexity, instead represents a truly fundamental component of the Universe. This idea dates back to antiquaty, being first proposed by the pre-Socratic philosophers (see, e.g., Thales of Miletus et al., 600BC), and being carried forward in the modern era by thinkers including Spinoza and Schopenhauer. Present-day philosophers including \cite{Goff2021-GOFPCF} have argued that the ``fundamental constituents of reality -- perhaps electrons and quarks -- have incredibly simple forms of experience''. Having forms of experience -- that there is something which it is like to `be' a fundamental particle -- would imply that these particles are necessarily endowed with SOULs (though of course it seems likely that the SOUL of a proton is less massive than the SOUL of a human.)

We return now to our previous estimate of the number of SOULs per star system, $10^{28} \leq N_S \leq 10^{33}$. The number of atoms in our Solar System is of the order $10^{57}$. Assuming the panpsychist interpretation to be correct, this suggests that each atom is endowed with $10^{-29} \leq N_S \leq 10^{-24}$ of a SOUL. We note that, given that the number of atoms in a human body is approximately $10^{28}$, this `atomic SOUL mass' range is in good agreement with the observation that humans are in possession of 1 SOUL each. This may be counted as a verified prediction of the Panpsychist Dark Matter (PDM) model.

\section{Conclusions and possibilities for further work}

We have put forward a novel solution to the observational `missing matter' problem. While this short preliminary work is only a first exploratory study, it is clear that this avenue has the potential to revolutionise both the cosmological understanding of dark matter and the Search for Extraterrestrial Intelligence. 

In addition to the above it should also be noted that the relatively high mass of the SOUL implies that the birth and death of an ensouled being emits a significant amount of gravitational waves. An exact treatment of the tensor perturbation produced by such events results in a particularly elegant description of the geometry of mortality, however, this description is too large to fit in this paper and is therefore beyond the scope of this work. Nevertheless, this implies that the future in the search for intelligent life may lie with the next generation of gravitational wave instruments rather than the more conventionally used optical/infrared telescopes.

\section*{ACKNOWLEDGEMENTS}
This work was supported by the cocktail menu at the Astronomer restaurant in Cambridge, without which the productive discussions which led to this paper would not have occurred.\\
Additionally, we extend our deepest gratitude to S. Gautama for generously financing our research on the dark matter candidate SOUL. This support not only made our work possible but also illuminated our path towards uncovering the mysteries of the universe

\section*{CONFLICT OF INTEREST}
The authors would like to declare no conflicts of interest. However, in case of the CTB or the PDM scenarios holding true, the possibility of the authors being influenced by these underlying fields cannot be discounted.

\bibliographystyle{abbrvnat}
\bibliography{Kiyaeva}

\end{document}

%% file: sao_cmd_author.tex
\def\squareforqed{\hbox{\rlap{$\sqcap$}$\sqcup$}}

\def\sq{\ifmmode\squareforqed\else{\unskip\nobreak\hfil
\penalty50\hskip1em\null\nobreak\hfil\squareforqed
\parfillskip=0pt\finalhyphendemerits=0\endgraf}\fi}

\def\utw{\smash{\rlap{\lower5pt\hbox{$\sim$}}}}

\def\udtw{\smash{\rlap{\lower6pt\hbox{$\approx$}}}}

\def\diameter{{\ifmmode\mathchoice
{\ooalign{\hfil\hbox{$\displaystyle/$}\hfil\crcr
{\hbox{$\displaystyle\mathchar"20D$}}}}
{\ooalign{\hfil\hbox{$\textstyle/$}\hfil\crcr
{\hbox{$\textstyle\mathchar"20D$}}}}
{\ooalign{\hfil\hbox{$\scriptstyle/$}\hfil\crcr
{\hbox{$\scriptstyle\mathchar"20D$}}}}
{\ooalign{\hfil\hbox{$\scriptscriptstyle/$}\hfil\crcr
{\hbox{$\scriptscriptstyle\mathchar"20D$}}}}
\else{\ooalign{\hfil/\hfil\crcr\mathhexbox20D}}%
\fi}}